\documentclass[11pt,letterpaper,twocolumn]{article}

\usepackage[11pt,nocopyright]{sigmin}

\usepackage{silence}
\usepackage[hyphens]{url}
\usepackage[breaklinks,colorlinks]{hyperref}
\usepackage[usenames,dvipsnames]{xcolor}
\hypersetup{citecolor=blue,linkcolor=blue}
\usepackage{amsmath,amsopn,amssymb}
\usepackage{subfigure}
\usepackage{endnotes,microtype,xspace,graphicx,fancyvrb,multirow}
\usepackage{booktabs}
\usepackage{array,underscore,relsize}
\usepackage[T1]{fontenc}
\usepackage{times}
\usepackage{fancyhdr,lastpage}
\usepackage{enumitem}
\usepackage[labelfont=bf,font=small,skip=5pt]{caption}
\usepackage{fp}
\usepackage{siunitx}

\usepackage{balance}

\newcommand{\edger}{\cc{Edger8r}\xspace}
\newcommand{\ecall}{\cc{ECALL}\xspace}
\newcommand{\ocall}{\cc{OCALL}\xspace}

\newcommand{\cc}[1]{\mbox{\smaller[0.5]\texttt{#1}}}

\fvset{fontsize=\scriptsize,xleftmargin=8pt,numbers=left,numbersep=5pt}

\input{code/fmt}
\newcommand{\figrule}{\hrule width \hsize height .33pt}
\newcommand{\coderule}{\vspace{-1.6em}\figrule\vspace{0.8em}}

\def\Snospace~{\S{}}

\newcommand{\x}{$\times$\xspace}

\newif\ifdraft\drafttrue
\newif\ifnotes\notestrue
\ifdraft\else\notesfalse\fi

\input{glyphtounicode}
\newcolumntype{R}[1]{>{\raggedleft\let\newline\\\arraybackslash\hspace{0pt}}p{#1}}

\newcommand{\includepdf}[1]{
  \includegraphics[width=\columnwidth]{#1}
}

\newcommand{\squishlist}{
\begin{itemize}[noitemsep,nolistsep]
  \setlength{\itemsep}{-0pt}
}
\newcommand{\squishend}{
  \end{itemize}
}

\usepackage{tikz}

\usepackage{xstring}
\newcommand{\PP}[1]{
\vspace{2px}
\noindent{\bf \IfEndWith{#1}{.}{#1}{#1.}}
}

\gdef\therev{8dbda71}
\gdef\thedate{2017-10-24 18:26:08 -0400}

\begin{document}

\title{Leaking Uninitialized Secure Enclave Memory via Structure Padding \\(Extended Abstract)}

\ifdefined\DRAFT
 \pagestyle{fancyplain}
 \lhead{Rev.~\therev}
 \rhead{\thedate}
 \cfoot{\thepage\ of \pageref{LastPage}}
\fi


\author{
 Sangho Lee\;
 Taesoo Kim\;
\\\\
 \emph{Georgia Institute of Technology}
}

\date{}
\maketitle

\begin{abstract}
  Intel Software Guard Extensions (SGX) aims to provide
  an isolated execution environment, known as an enclave, for
  a user-level process to maximize its
  confidentiality and integrity.
  In this paper,
  we study how uninitialized data inside a secure enclave
  can be leaked via structure padding.
  We found that, during \ecall and \ocall,
  proxy functions that are automatically generated by
  the Intel SGX Software Development Kit (SDK) fully copy
  structure variables from an enclave to the normal memory
  to return the result of an \ecall function and
  to pass input parameters to an \ocall function.
  If the structure variables contain padding bytes,
  uninitialized enclave memory,
  which might contain confidential data like a private key,
  can be copied to
  the normal memory through the padding bytes.
  We also consider potential countermeasures against
  these security threats.
\end{abstract}

\section{Introduction}
\label{s:intro}


Intel Software Guard Extensions (SGX) is
a hardware-based trusted execution environment (TEE) technology that
allows a user-level process to have an isolated execution environment,
known as an \emph{enclave}.
Even system software, such as operating system and hypervisor,
cannot access enclave memory,
known as the enclave page cache (EPC), because
Intel processor's memory management unit (MMU)
prohibits such attempts at hardware level.
Further,
physical attacks, such as a cold-boot attack,
are not possible because every data stored in EPC banks
is encrypted by the memory encryption engine (MEE) before
it leaves from the processor package.
%


Intel SGX has two restricted interfaces to allow
SGX encalves to interact with non-enclave applications:
\ecall and \ocall.
Non-enclave applications should use the \ecall interfaces to
execute trusted functions within enclaves.
They cannot call any other enclave functions without
corresponding \ecall interfaces.
Also, 
SGX enclaves should use the \ocall interfaces to
execute untrusted functions (e.g., system calls).
Their any other attempts to execute untrusted functions
(e.g., jumping into non-enclave code) result in
faults.

Intel SGX Software Development Kit (SDK)
is shipped with a tool called \edger~\cite{intel:edger8r} that
automatically and securely generated code for \ecall and \ocall interfaces.
Although
SGX enclaves can access both EPCs and normal memory,
non-enclave applications can only access the normal memory.
Thus,
all input and output values for the \ecall and \ocall interfaces between them
need to be stored in the normal memory first and then
copied to the memory of callee and caller later.
%
%
The \edger tool creates all such edge routines automatically.
It decodes the user-provided enclave definition language (EDL) files specifying
\ecall and \ocall interfaces, and
generates \emph{proxy functions} to securely exchange input and output parameters for
the interfaces.
That is, the proxy functions
copy data between enclaves and non-enclave applications
as well as check the sanity of input or output data
(e.g., they check the address range if a parameter is a pointer).

The proxy functions generated by \edger need to be secure because
they are designed to copy certain data
(i.e., input values to an \ocall interface and
a return value of an \ecall interface)
from the enclave to the normal memory.
If the proxy functions have security problems,
they might be exploited to extract sensitive data from the enclave that
is neither input to \ocall nor output of \ecall,
which results in incomplete confidentiality of Intel SGX.

In this paper,
we explore the security problems of the proxy functions
generated by the \edger tool for
the \ecall and \ocall interfaces of Intel SGX.
More specifically,
we focus on the possibility of data leakage because of \emph{structure padding}.
When handling structure data types (e.g., \cc{struct} in C),
modern compilers intentionally align
their members
by putting some padding bytes,
perhaps to reduce memory/cache access time~\cite{raymond:packing}.
These padding bytes, however, are usually ignored
when initializing structure variables such that
they can contain uninitialized memory values~\cite{lu:unisan}.
If the proxy functions generated by \edger
do not consider this security problem,
uninitialized enclave data can be leaked through padding.

We confirm that
when the data types of
input values to an \ocall interface or
a return value of an \ecall interface are
structures containing padding bytes,
uninitialized enclave data is copied to
the normal memory via the padding bytes
during \ocall or \ecall by
the proxy functions.
This is because
the proxy functions are generated to
copy the entire memory of a structure variable,
not to copy its individual members.
That is,
they do not perform deep copy.

We expect that
the impact of this data leakage problem through uninitialized structure padding is
similar to that of the Heartbleed vulnerability.
As Heartbleed does,
this security problem allows us to leak
a number of bytes from secure enclaves.
More importantly,
all data within secure enclaves is supposed to be in \emph{plaintext}.
This is because
the secure enclaves, by design, ensure the data confidentiality and isolation
such that we do not need to redundantly encrypt the data.
That is,
the secure enclaves likely manage
the plaintext of sensitive information
(e.g., RSA private key, password, and biometric information)
in their memory, which can be leaked through
the uninitialized structure padding.
Therefore,
when developing SGX applications,
developers should carefully consider whether
their applications can suffer from this critical security problem.

Possible countermeasures to
this uninitialized padding problem are as follows:
(1) perform per-member deep copy during \ecall and \ocall,
(2) use the \cc{\#pragma pack} directive to avoid padding,
(3) enforce \cc{memset} to fully initialize structure variables, and
(4) adopt advanced structure initialization techniques~\cite{lu:unisan,milburn:safeinit}.

\section{Background: Structure with Padding}
\label{s:background}

\begin{figure}[t]
  \input{code/struct.c.tex}
  \vspace{1em}
  \coderule
  \caption{Structure with padding (in \cc{x86_64}).
  The total size of this structure is 24 bytes because of the padding.}
  \label{c:struct}
\end{figure}

\autoref{c:struct} shows an example C structure, \cc{test_struct}, which
contains padding bytes used for aligning its member variables.
The \cc{test_struct} structure has three member variables,
\cc{val1}, \cc{val2} and \cc{val3}, in which
the first and third member variables' sizes are eight bytes whereas
the second member variable's size is one byte.
If there is no \cc{\#pragma pack} directive,
modern C compilers will put seven-byte padding between
\cc{val2} and \cc{val3} to align all of the three member variables
for memory access efficiency~\cite{raymond:packing}, so that
the size of \cc{test_struct} will be 24 bytes.
That is,
if we have a variable \cc{ts} whose type is \cc{test_struct},
initializing its individual members
(i.e., \cc{ts.val1=0}, \cc{ts.val2=0}, \cc{ts.val3=0})
is not enough to fully clean up this variable such that
its padding bytes can contain uninitialized data.
Instead,
we have to
explicitly initialize padding bytes by using
\cc{memset(\&ts, 0, sizeof(test_struct))}.

\section{Uninitialized Enclave Memory Leakage via Padding}
\label{s:enclave-padding}

In this section,
we explain how uninitialized enclave memory can be leaked
through structure padding.
We focus on the following two cases:
(1) \ecall returning a structure variable with padding (\autoref{ss:ecall-padding}) and
(2) \ocall having an input structure variable with padding (\autoref{ss:ocall-padding}).
We used
Intel SGX SDK for Linux version 1.9 and
a real system with Core i7-6700K to test the explained problems
and confirmed that all the problems really existed.

\subsection{\ecall Returning Padded Structure}
\label{ss:ecall-padding}
\begin{figure}[t]
  \input{code/ecall-struct.c.tex}
  \vspace{1em}
  \coderule
  \caption{\ecall returning \cc{test_struct}.
  The padding inside \cc{test_struct} is not initialized.}
  \label{c:ecall-struct}
\end{figure}
We explain how \ecall functions returning a structure variable
can leak sensitive enclave data through padding.
\autoref{c:ecall-struct} shows an example \ecall function that
returns a padded structure, \cc{test_struct},
explained in \autoref{s:background}.
This function
receives two input values,
performs some computations with them, and
eventually returns \cc{test_struct}.
Since
this function does not clear the padding bytes of \cc{test_struct},
they can contain uninitialized enclave data and be returned to
a proxy function generated by the \edger tool.

\begin{figure}[t]
  \input{code/enclave-t.c.tex}
  \vspace{1em}
  \coderule
  \caption{Proxy function for \cc{ecall_test_struct}.
    The entire content of \cc{test_struct} including
  the padding is returned to the non-enclave memory.}
  \label{c:proxy-ecall}
\end{figure}
\autoref{c:proxy-ecall} represents a proxy function for
the \cc{ecall_test_struct} function, which is
automatically generated by \edger to be executed inside an enclave.
As shown in Line 16,
this proxy function just fully copies
the return value of \cc{ecall_test_struct} into
\cc{ms->ms_retval},
a non-enclave marshalled structure for storing
input and out values for this \ecall function,
instead of individually copying its members to \cc{pms->ms_retval}.

Therefore,
we conclude that
an \ecall proxy function can copy uninitialized padding bytes from
an enclave to a non-enclave memory region when returning a structure variable,
which can potentially contain uninitialized sensitive enclave data.

\subsection{\ocall Receiving Padded Structure as Input}
\label{ss:ocall-padding}
\begin{figure}[t]
  \input{code/ocall-struct.c.tex}
  \vspace{1em}
  \coderule
  \caption{\ocall receiving \cc{test_struct} as input.
  This function can access the data in the padding if it is not initialized.}
  \label{c:ocall-struct}
\end{figure}
\begin{figure}[t]
  \input{code/enclave-t2.c.tex}
  \vspace{1em}
  \coderule
  \caption{Proxy function for \cc{ocall_test_struct}.
    The entire content of \cc{test_struct} including the padding
  is passed to \cc{ocall_test_struct()} as an input parameter.}
  \label{c:proxy-ocall}
\end{figure}
Next,
we explain how \ocall functions receiving a structure variable as input
can leak sensitive data through padding.
\autoref{c:ocall-struct} shows an example \ocall function, \cc{ocall_test_struct}, that
receives a padded structure, \cc{test_struct}, as input.
This function tries to access every single byte of
a \cc{test_struct} variable, \cc{ts},
implying that if this input struct contains padding with
sensitive enclave data, this function can access them it also.

\autoref{c:proxy-ocall} shows a proxy function of \cc{ocall_test_struct} that
will be executed inside an enclave.
This function allocates a non-enclave memory region for a marshalled structure (Line 16) and
copies an input structure variable, \cc{ts}, into it (Line 25)
while doing not handle the individual members of the structure.

Therefore,
we conclude that
an \ocall proxy function can copy uninitialized padding bytes from
an enclave to a non-enclave memory region when
the corresponding \ocall function receives a structure variable as input,
which can potentially contain uninitialized sensitive enclave data.

\section{Potential Countermeasures}
\label{s:discuss}
In this section,
we explain possible countermeasures against
the sensitive enclave data leakage problem because of padding,
which can be implemented in the future.
First,
we can revise the \edger tool to generate proxy functions that
individually copy the member of a structure variable during
\ecall and \ocall (i.e., deep copy).
For example,
in \autoref{c:proxy-ocall},
we can do
\cc{ms->ms_ts.val1=ts.val1}, 
\cc{ms->ms_ts.val2=ts.val2}, and
\cc{ms->ms_ts.val3=ts.val3}, instead of
\cc{ms->ms_ts=ts} to do not leak uninitialized padding.
However,
it makes proxy functions be complicated especially when
they are dealing with a complex structure variable recursively containing
other structures.
Second,
we can enforce \cc{\#pragma pack} directive to eliminate any padding,
but it introduces performance overhead because of a lack of memory alignment.
Third,
whenever we allocate a structure variable inside an enclave,
we can use \cc{memset()} to fully initialize all of its memory.
This countermeasure can avoid any potential data leakage problems,
but it can experience performance degradation.
Fourth,
as Lu et al.~\cite{lu:unisan} and Milburn et al.~\cite{milburn:safeinit} do,
we can implement advanced memory initialization techniques to selectively
initialize padding bytes only when they can leak sensitive data.
We believe that
this is the most practical solution against the explored problems such that,
in the future, we will figure out the best memory initialization technique
for Intel SGX to eradicate the uninitialized enclave data leakage problem.
\section*{Responsible Disclosure}
\label{s:disclose}
We have reported this uninitialized padding problem to Intel on June 23, 2017.
After having numerous discussions with us,
Intel SGX SDK developers informed that
this padding issue will be explicitly warned
in the SDK documentation to describe this issue
and potential mitigation approaches.

\balance
\bibliographystyle{acm}
\footnotesize
\bibliography{p,sslab,conf}
\end{document}